\documentclass[epjCONF]{svjour}
\usepackage{graphics}
\usepackage[varg]{txfonts} 
\usepackage[latin1]{inputenc}
\session-title{Conference Title, to be filled}
\begin{document}
\title{Refined reconstruction and calibration of the missing transverse energy in the ATLAS detector}
\author{Marianna Testa \thanks{\email{marianna.testa@lnf.infn.it}} on behalf of the ATLAS collaboration }
\institute{LNF-INFN }
\def\MET{\ensuremath{E_{\mathrm{T}}^{\mathrm{miss}}}\,}
\def\etmiss{$\textbf{\textit{E}}_{\rm T}^{\rm miss}$}
\def\elmiss{$\textbf{\textit{E}}_{\rm T}^{\rm miss}$}
\def\pT{$p_{\mathrm{T}}$}
\def\abseta{\ensuremath{|\eta|}}
\def\Zll{\ensuremath{Z \rightarrow \ell\ell}}
\def\Wln{\ensuremath{W \rightarrow \ell\nu}}
\def\Zmm{\ensuremath{Z \rightarrow \mu\mu}}
\def\Zee{\ensuremath{Z \rightarrow ee}}
\def\myFigSize{5.7cm}
\def\px {$E_{x}^{\mathrm{miss}}$}
\def\py {$E_{y}^{\mathrm{miss}}$}\def\phimiss{$\phi^{\rm miss}$}
\def\sumet{$\sum E_{\mathrm{T}}$}
\newcommand{\emiss}[1]{\ensuremath{E_{#1}^{\mathrm{miss}}}}
\newcommand{\etreso}{\ensuremath{\sigma(\emiss{x},\emiss{y})}}
\def\ifb{\mbox{fb$^{-1}$}}
\def\ipb{\mbox{pb$^{-1}$}}
\def\inb{\mbox{nb$^{-1}$}}
\def\etmissmag{$E_{\mathrm{T}}^{\mathrm{miss}}$}
\def\elmissmag{$E_{\mathrm{T,Z}}^{\mathrm{miss}}$}
\def\Wmn{\ensuremath{W \rightarrow \mu\nu}}
\def\Wen{\ensuremath{W \rightarrow e\nu}}
\def\pTZ{$p_{\mathrm{T}}^Z$}
\def\rts {\ensuremath{\sqrt{s}}}
\abstract{
The measurement of the missing transverse energy (\etmiss)  is fundamental  for many analyses at LHC. Good \etmiss\ resolution and calibration  are essential for searches of new physics as well as precise measurements. We describe a refined reconstruction and calibration of \etmiss\ developed by ATLAS and its performances on events containing Z and W bosons. The data sample was collected in proton-proton collisions at a center-of-mass energy of 7 TeV, and corresponds to an integrated luminosity of about 36 \ipb.  The determination of the absolute scale of the \etmiss, fundamental for determining systematic uncertainties in all analysis involving \etmiss\ measurements, is also presented.
}

\maketitle

\section{\etmiss\ reconstruction}
The \etmiss\ reconstruction includes contributions from energy deposits in the calorimeters and muons reconstructed in the muon spectrometer.  The two \etmiss~ components are calculated as: $        E_{x(y)}^{\mathrm{miss}} = E_{x(y)}^{\mathrm{miss,calo}}
                                   + E_{x(y)}^{\mathrm{miss,\mu}}.$ The \etmiss\ calorimeter term $E_{x(y)}^{\mathrm{miss,calo}}$ is reconstructed using  calorimeter cells calibrated according to the 
reconstructed physics object to which they belong.
 Calorimeter cells are associated with a
 reconstructed and identified high-\pT\  parent object in a chosen order:
electrons, photons, hadronically decaying $\tau$-leptons,
jets and muons. 
 Cells not associated with any such objects are also taken into account in the \etmiss~ calculation. Their contribution, named  $E_{T}^{\mathrm{miss,CellOut}}$ hereafter,  is important for the \etmiss~ resolution. 
Once the cells are associated with objects as described
above, the \etmiss \ calorimeter term is calculated as follows~\cite{metpaper}:
\begin{eqnarray}
      E_{x(y)}^{\mathrm{miss,calo}} =
              E_{x(y)}^{\mathrm{miss},e}         +
              E_{x(y)}^{\mathrm{miss},\gamma}    +
              E_{x(y)}^{\mathrm{miss},\tau}      +
              E_{x(y)}^{\mathrm{miss,jets}}    
 +
              E_{x(y)}^{\mathrm{miss,softjets}}   + 
              E_{x(y)}^{\mathrm{miss,calo},\mu} +
              E_{x(y)}^{\mathrm{miss,CellOut}}
\label{eq7} 
\end{eqnarray}

where each term
is calculated from the negative sum of calibrated cell energies inside the corresponding objects, as: $ E_{x}^{\mathrm{miss,term}}=-\sum_{i=1}^{N_{\rm cell}^{\rm term}}E_i\sin\theta_i\cos\phi_i ,\nonumber 
 E_{y}^{\mathrm{miss,term}}=-\sum_{i=1}^{N_{\rm cell}^{\rm term}}E_i\sin\theta_i \sin\phi_i\ $
where $E_i$, $\theta_i$ and $\phi_i$ are the energy, the polar angle
and the azimuthal angle, respectively.
The various terms in Equation \ref{eq7} are described in the following:      
\begin{itemize}
\item
              $E_{x(y)}^{\mathrm{miss},e}$, 
              $E_{x(y)}^{\mathrm{miss},\gamma}$,
              $E_{x(y)}^{\mathrm{miss},\tau}$   are reconstructed from cells in clusters associated to  electrons, photons and $\tau$-jets from hadronically decaying $\tau$-leptons, respectively;
\item
 $E_{x(y)}^{\mathrm{miss,jets}}$ is reconstructed from cells in clusters associated to calibrated  jets with \pT~$>$ 20 GeV;
\item
 $E_{x(y)}^{\mathrm{miss,softjets}}$ is reconstructed from cells in clusters associated to jets with 
 7 GeV $<$~ \pT~$<$ 20 GeV;
 \item
 $E_{x(y)}^{\mathrm{miss,calo},\mu}$ is the contribution to \etmiss\
          originating from the energy lost by muons in the calorimeter
 \item
 the $E_{x(y)}^{\mathrm{miss,CellOut}}$ term is
calculated from the cells in topoclusters which are not included in the
reconstructed objects. Low-\pT~ tracks are also used  to recover low \pT~ particles not reaching  the calorimeters. Furthermore the track momentum is used instead of the topocluster energy for tracks associated to topoclusters,  thus exploiting the better calibration and resolution of tracks at   low momentum compared to topoclusters
\end{itemize}
In order to suppress noise contribution, only cells belonging to three-dimensional topological clusters~\cite{clusters} are used.
The \etmiss\ muon term is calculated from the momenta of muon tracks reconstructed with  $\abseta<2.7$: $       E_{x(y)}^{\mathrm{miss},\mu} =  - \sum_{\mathrm{muons}} p_{x(y)}^{\mu}$
where the summation is over selected muons.
If the muon is isolated (not nearby a reconstructed jet),  the  energy  
lost by the muon in the calorimeters is not added to the calorimeter  term to avoid double counting of energy. Otherwise the muon spectrometer measurement of the muon momentum after correcting for energy loss in  
the calorimeter is used ($E_{x(y)}^{\mathrm{miss,calo},\mu}$ contribution)

\section{Study of \etmiss~ performance}
\label{sec:2}
During  2010 a large number of proton-proton collisions, at a centre-of-mass energy of 7 TeV were recorded.  Approximately 600 \inb~ for 
 jet events, 0.3 \inb~ for minimum bias events and 36~\ipb for \Zll~ and \Wln~ channels have been  used  to check both   agreements on \etmiss\ distributions between data and MonteCarlo (MC) simulation  and to study  the  performances of the \etmiss\ reconstruction   in terms of resolution and scale. Details on this study  can be found in Ref.~\cite{metpaper}.  In particular in minimum bias, di-jet and  \Zll~ events no true \etmiss\ is  expected and resolution can be estimated as the width of distribution of $\emiss{x}$ and $\emiss{y}$.  In fig.~\ref{fig:resol_tutti} (left) the resolution from data at $\sqrt{s}=7$ TeV for  \Zll~ events, minimum bias and  di-jet events
as a function of the total transverse energy in the event, obtained by summing the \pT~ of muons and the \sumet~ in calorimeters. In fig.~\ref{fig:resol_tutti} (right) the \etmiss~ resolution is shown for MC events also for  \Wln~ MC events. The resolution of the two \etmiss~ components is fitted with a function $\sigma= k \cdot \sqrt{\Sigma E_{\mathrm{T}}}$. A good  agreement in the resolution is found between data and MonteCarlo simulation. 
In order to study the \etmiss scale it's useful to consider  the component of \etmiss~ along the  $Z$ direction (\elmissmag) which is  sensitive to the balance between
the leptons and the hadronic recoil. 
Figure  \ref{figure_Diagnostic}  shows the mean value of  \elmissmag\  as a
function of \pTZ. The  negative bias for low values of \pTZ\ is due to  the 
underestimation of magnitude of the hadronic recoil  dominated by $E_{\mathrm{T}}^{\mathrm{miss,CellOut}}$ and by softjets. 
\begin{figure}
\resizebox{0.45\columnwidth}{!}{\includegraphics{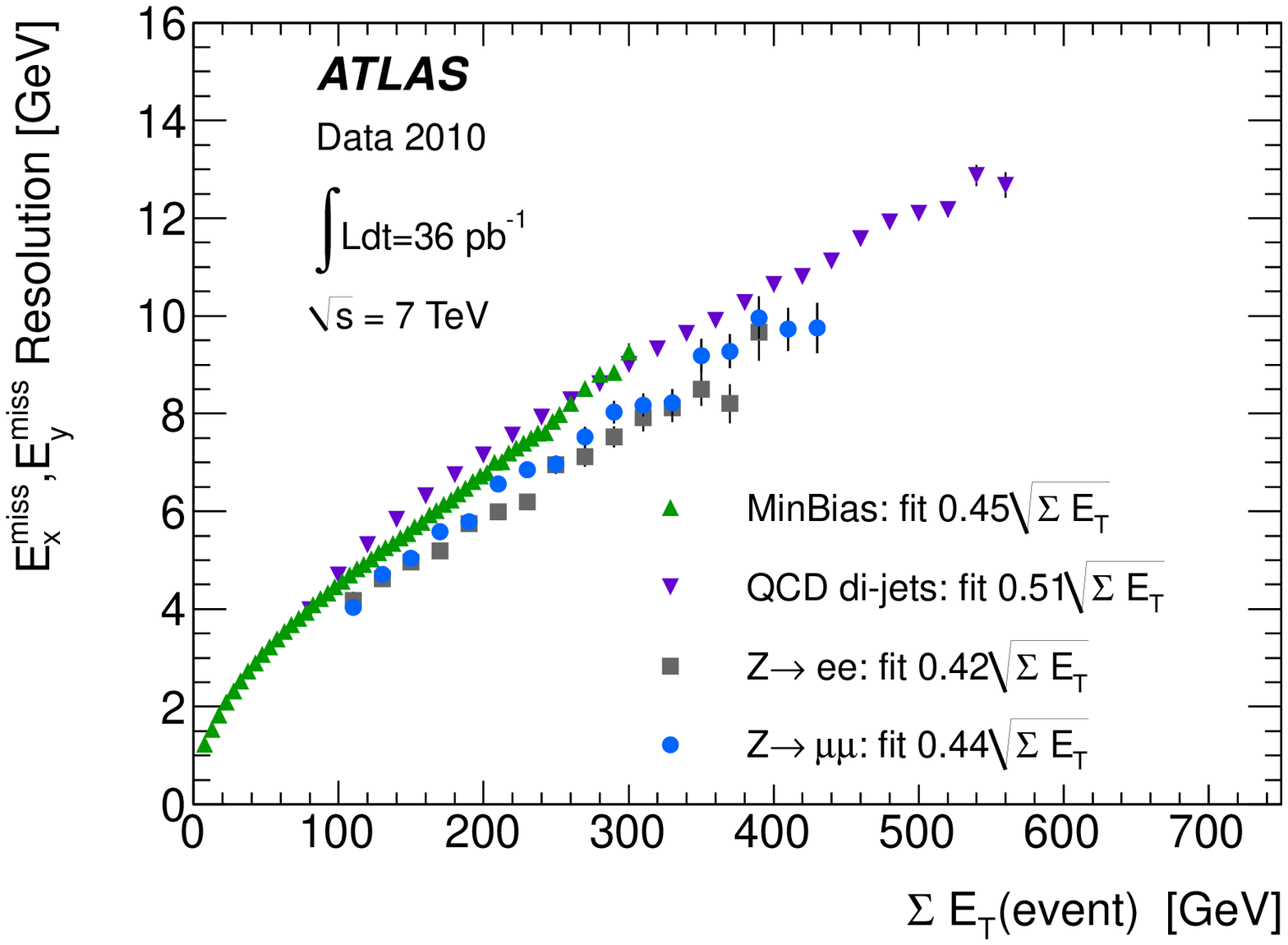}}
\resizebox{0.45\columnwidth}{!}{\includegraphics{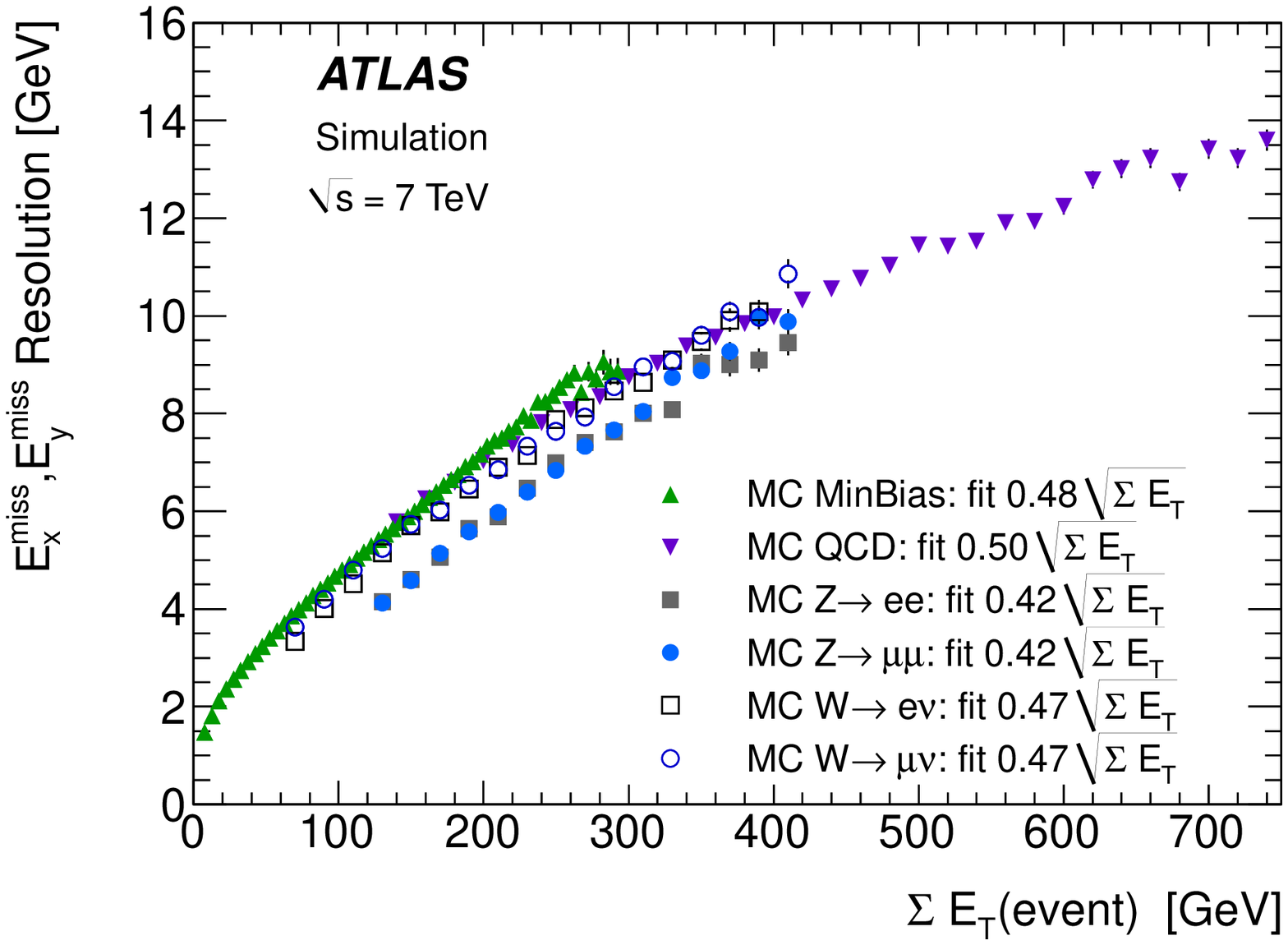}}
\caption{ \px~and \py~ resolution  as a function of 
the total transverse energy in the event calculated by summing the \pT~ of muons and the total transverse energy in the calorimeter in data  at  $\sqrt{s}$ = 7 TeV (left) and MC (right). 
}
\label{fig:resol_tutti}
\end{figure}

\begin{figure}
\resizebox{0.45\columnwidth}{!}{\includegraphics{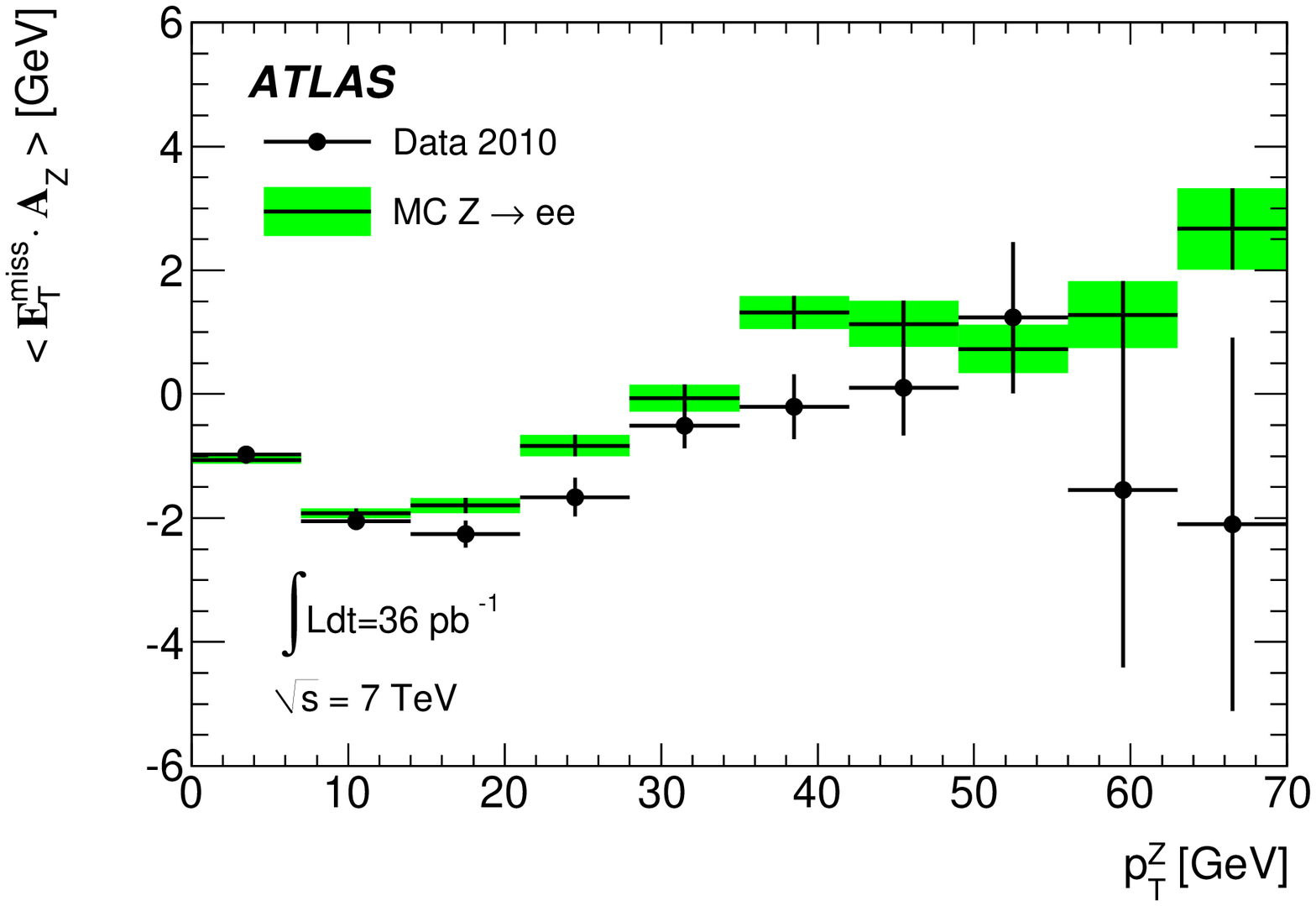}}
\resizebox{0.45\columnwidth}{!}{\includegraphics{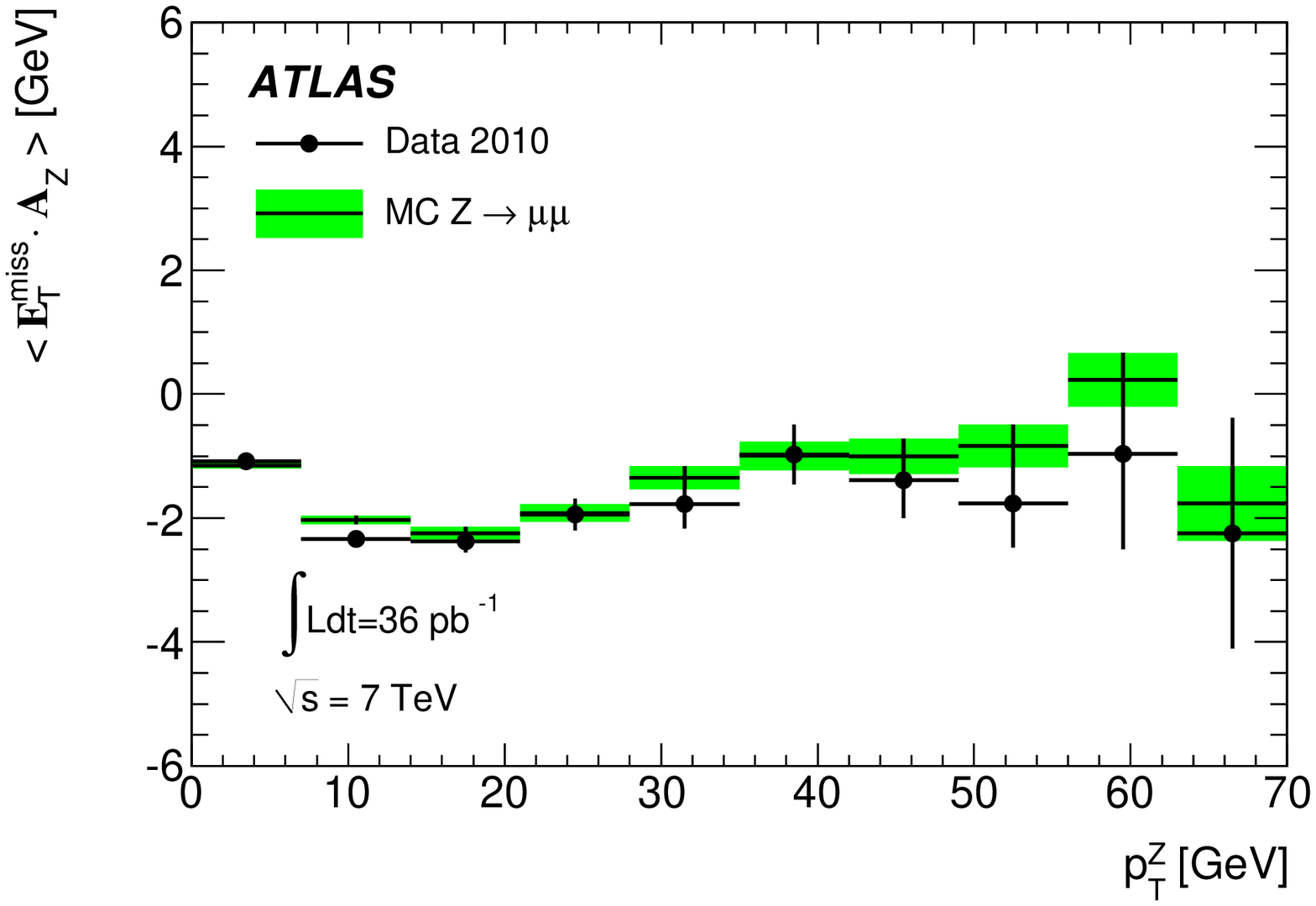}}
\caption{ Mean values of \elmissmag\ 
 as a function of \pTZ~ in $\Zee$ (left) and $\Zmm$ (right) events}
\label{figure_Diagnostic}
\end{figure}
\subsection{Evaluation of the systematic uncertainty on the \etmiss~ scale}
The knowledge of  the systematic uncertainty on the \etmissmag~ scale  is fundamental for any analysis involving \etmiss\ measurements.
The  overall systematic uncertainty on the \etmissmag~ scale 
can be calculated from the uncertainty on each high \pT\ reconstructed object and from the uncertainty on softjets and CellOut term (Equation \ref{eq7}) , which are evaluated to be ~13\% and ~10 \% respectively.  In \Wln\ events the overall \etmiss\ systematics uncertainty  is on average 2.6\% for both electron and muon channel.

\section{Determination of the \etmiss~ scale from $\textit{{\textbf{W}}} \  \rightarrow \ell \nu$  events}
The  determination of the  absolute 
\etmissmag~ scale  is important 
in a range of  analyses  involving \etmissmag\ measurements,
ranging from precision measurements to searches  for  new  physics.
\etmissmag\ scale has been determined from data with two methods~\cite{metpaper}. The
first uses a fit to the distribution of transverse mass, $mT$, of the lepton \etmiss\ system (fig.~\ref{fig:met_scale}, left). The second uses the dependence between the neutrino and lepton momenta (fig.~\ref{fig:met_scale}, right). Both methods have consistent results  and  find good  agreement   between   data  and   MC simulation  for   the \etmissmag\ scale. The  uncertainty on the scale is  about 2\% with 36 \ipb.

\begin{figure}
\resizebox{0.45\columnwidth}{!}{\includegraphics{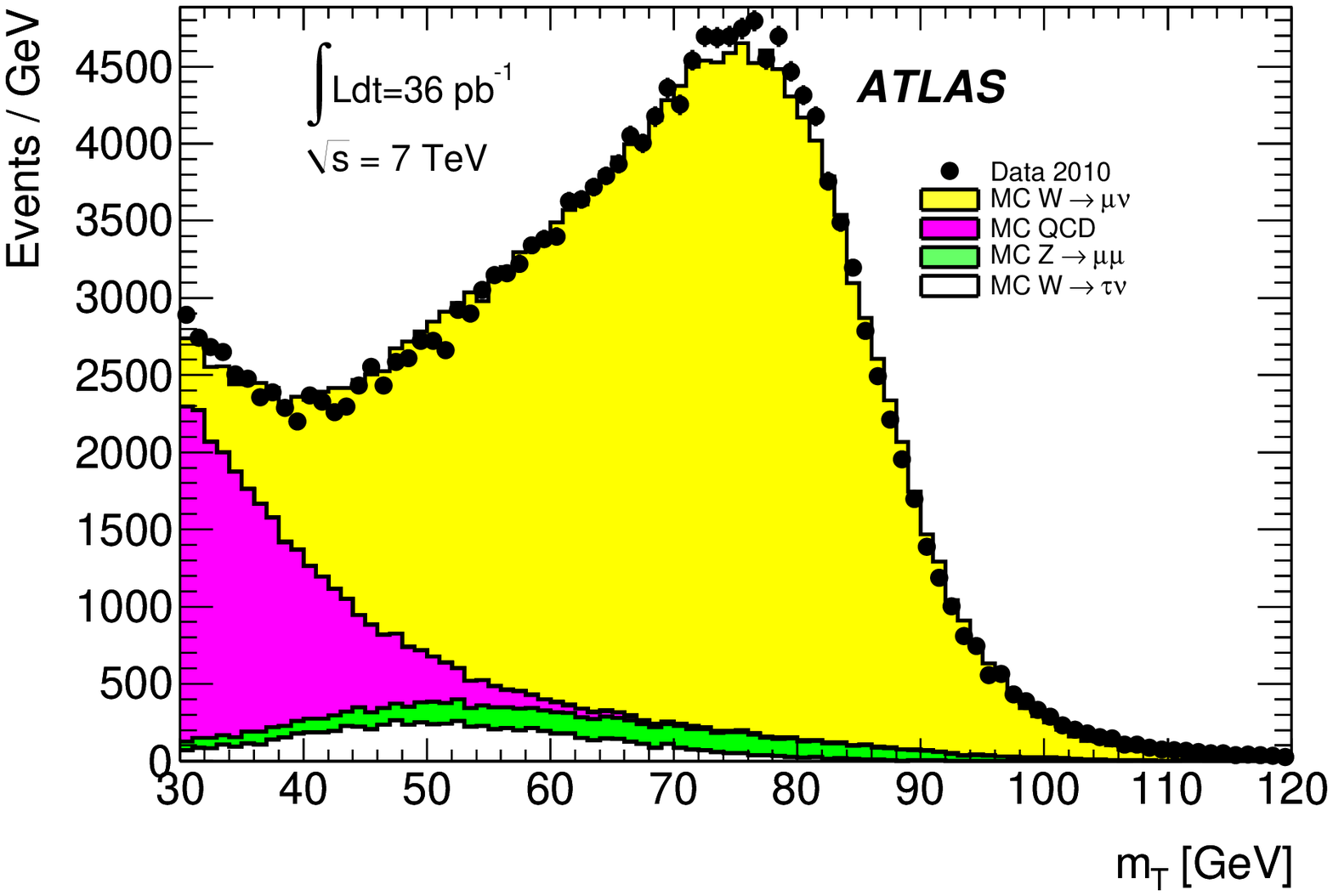}}
\resizebox{0.45\columnwidth}{!}{\includegraphics{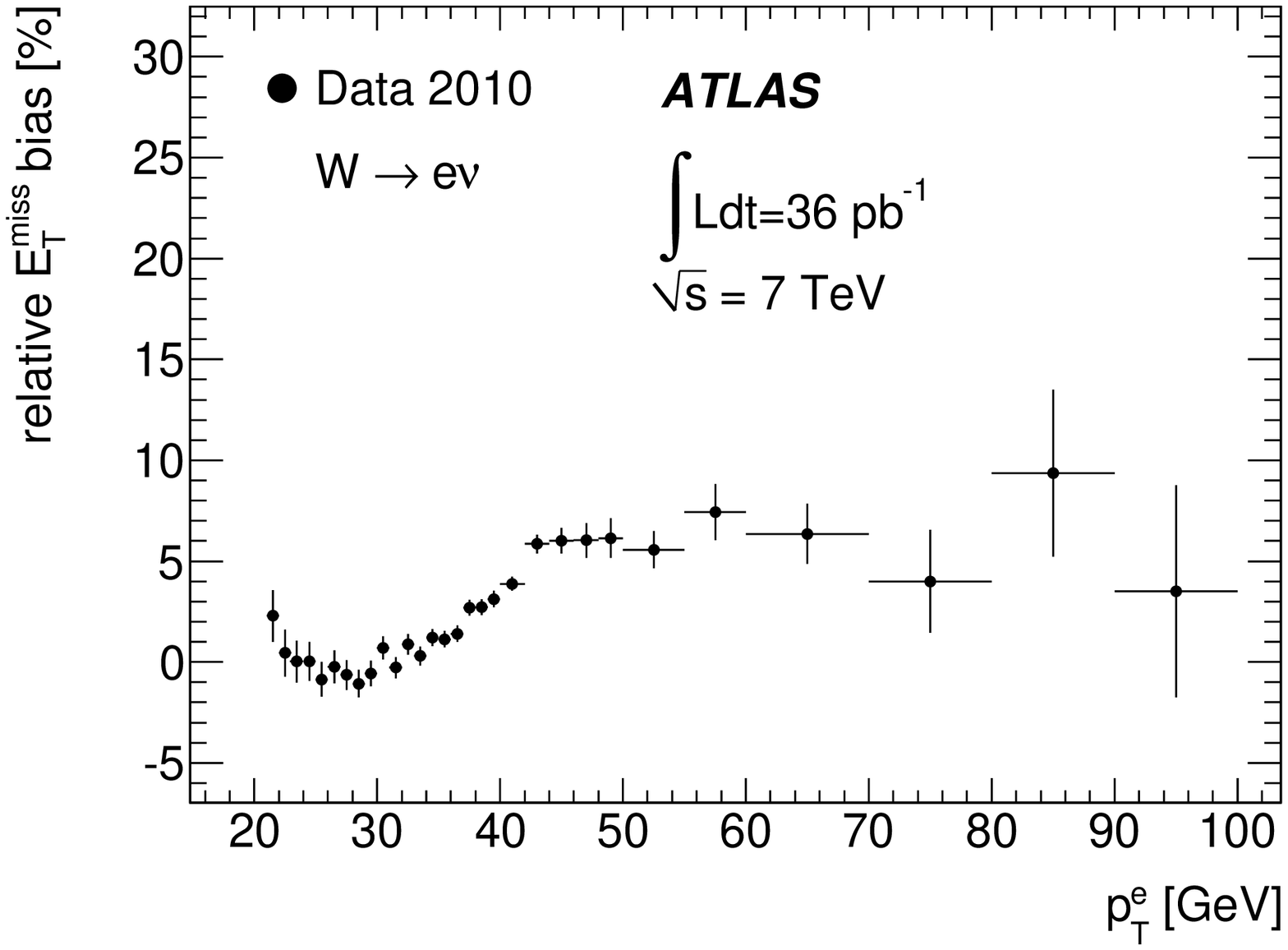}}
%
\caption{
Left: Fit to $m_T$ distribution  for \Wmn\ events. Right:                        Relative bias,  $\langle$\etmissmag$\rangle - \langle$\etmissmag$^{\rm ,True}\rangle$)/$\langle$\etmissmag$^{\rm ,True}\rangle$, in the reconstructed \etmiss\ for \Wen\ events.
}
\label{fig:met_scale}
\end{figure}

\end{document}